# Practitioners At The Limit

## Bereavement, Mockery and Ideology in Response to Crisis

Shahpour Akhavi
Science and Technology Studies
Graduate Program
York University
Toronto ON Canada
sakhavi@yorku.ca

## ABSTRACT

After decades in which the software industry heroized its technical employees, our current moment finds those employees in crisis. Economically, they are squeezed by a job market that has turned on them since Elon Musk gutted Twitter in 2022. Politically, their craft is continually pressed into the service of an ever tighter alliance between Big Tech and authoritarianism. Professionally, they find themselves less and less able to contribute anything good to anyone within business models that are running out of room to pretend they do anything but extract. Technically, they are confronted with a much-hyped technology, generative AI, that distorts their work while purporting to make them redundant. And emotionally, they are simply not ok — as designer and developer Andrew Sempere puts it, "I think the word I'm looking for is: *bereft*."

This study draws from a series of practitioner interviews undertaken for a current dissertation in STS and from ongoing content analysis of practitioners' public statements on blogs, microblogs, podcasts and videos to argue that much of the workforce of technocapitalism is actively unaligning itself from that system's priorities and seeking new ones. Countercultural responses among practitioners are proliferating, including mockery of technocapitalists and technocapitalism, rehabilitation of Luddism, interest in organized labour, and endorsement of alternate systems including anarchism and communism.

Whether software practitioners will form an organized enough response to decisively influence the crises that beset them (and, often through their products, the wider world) remains to be seen. But heightened critical attention should be paid to this centrally-located population of technosocial actors in their new and volatile state.

## 1 Introduction

Software is catalyzing the polycrisis. Despite its clean reputation compared to other industries, software energy use was considerable [27] even prior to the skyrocketing resource demands and pollutive effects of generative AI [34]. Digital behaviour modification and rage-bait business models have poisoned civic discourse and fostered the rise of reactionary and fascist movements. [43, 63] Complementarily, authoritarian powers use software to conduct oppressive surveillance and disrupt popular resistance [17, 64]. The world's wealthiest individuals, many of whom are allies and / or partners of authoritarian state regimes [40], head software companies that often depend on poorly paid and often gruelling invisible labour [78].

This is not the future that software practitioners — the designers, developers, agilists and others who collaboratively create and maintain the world's software products — were promised. In 1998, when my own software career began in San Jose, California, it was a New Economy, a post-scarcity economy in which stock options actually did make a fair few practitioners into millionaires overnight. Federal Reserve Chair Alan Greenspan avowed it was 'like discovering a new planet.' [77] But despite the E*TRADE billboards flanking US-101 in the Valley ("Making money is only half the fun… No wait, that's pretty much it"), many felt the most rewarding aspect of the field was its positive impact on others' lives. The graphical user interface we coded today saved real people hours of labour tomorrow; new design technologies were bringing elegance and verve online; if nothing else, we were saving trees. Of course, there were bugs, but those were fascinating puzzles to be solved by what our employers assured us in all-hands meetings were our superior capacities. Those same capacities merited, they said, the perks they lavished — free cafes, flexible hours, individual offices with doors, and foosball tables (even if we were too busy to play). And as they professed belief in us, so we often did in them: Many of my colleagues fanboyed (being mostly boys) over Steve Jobs, dreamed of their own IPOs, and read *Fast Company* as avidly as Webmonkey.com.

But in 2026, scarcity is back. The billboard, in retrospect, may have been right: Making money was the only point. I have not seen a front-line software practitioner with a closed-door office since 2004. Return-to-office mandates [72], AI use mandates [55], and never-ending layoffs [32, 53, 62] clarify the limits of managers' independent affection for their technical staff. But more than this, after decades of platform capitalism [68] and algorithmic injustice [e.g. 10, 67], practitioners are questioning how much good (and even how much harm) their technical efforts are doing for and to their fellow human beings.

This paper draws from practitioner interviews and online public statements to assess the impact of the current confluence of negative pressures on practitioners' alignment with the technocapitalist forces that direct their industry. I argue that for many, this alignment is now at a breaking point, as attested by proliferating expressions of dissent. I will look at three categories of such dissent in particular: 1) *Bereavement*, in which practitioners publicly express anguish, despair, grief, and anger about their individual and collective predicament; 2) *Mockery*, in which they use humour to target capitalist figures and ideologies that they hold responsible for that predicament; and 3) *Endorsement* of competitor ideologies to technocapitalism, such as communism, anarchism, Luddism, and labour collectivism.

Ten interviews conducted to date for my current dissertation study, of an eventual thirty, were analyzed for this paper. Subjects were selected from my own network of ex-colleagues and from those with whom I have connected via social media based on their public commentary on software practice. All hold professional titles in software development, UX (user experience) design, and / or agile process management. Semistructured interviews took place during 12/2025-3/2026 via Zoom webconferencing software, with video, audio and auto-transcription captured by the software. Additionally, I draw on a collection of public practitioner online statements in blogs, microblogs, podcasts, and videos. Systematic collection of these statements began in 2021 based on contemporaneous popular lists of "practitioners worth following" in each field, was incrementally expanded based on others' online interactions with the original set, and remains ongoing. Sources in this archive are identified as practitioners via job title (either stated in social media bios or located on LinkedIn) and / or descriptions of their current work (e.g. references to their code repositories). In both methods, a diversity of practitioner gender, ethnicity, and national origin was sought, but no claim is made regarding sample represenationality of the global practitioner population in these regards, and all data was collected in English. Interviewees were all located in North America and Europe at the time of their interviews, while online comment also comes from practitioners located in Asia and the South Pacific. Practitioners as a whole are engaged and voluble about their practice and its context, and those contacted were generally eager to participate, with a large majority waiving anonymity regarding their comments; however, in this paper, interviewees have been named only by job title.

A minor investigation, for the limited purpose of assessing the deployment of Luddism on the professional networking site LinkedIn was also undertaken via social media scraping methods, described in the relevant section below.

The following sections detail findings related to bereavement, mockery and ideological endorsement. Subsequently, a short conclusion speculates on the implications of these responses for effective action that can resist, moderate or reform their causes.

## 2 Hitting Limits: Practitioner Bereavement

In a February 2026 blog post titled "Don't Die Lil Thing," software designer and developer Andrew Sempere says,

> It's not the internet that's dead, it's the entire tech industry and it wasn't an accident, it was a murder to try and cash in on the life insurance money.... I think the word I'm looking for is: *bereft*. [66]

Danilo Campos, another designer and developer, elaborates, "The power of technology has been coopted, and that which once brought us joy has been contaminated altogether. This is discouraging and traumatic." [16]

Sempere and Campos are hardly alone among software practitioners in identifying strong emotion about the pass in which the field finds itself. Dismayed by Microsoft's use of nonsensical "AI slop" illustration for its technical documentation [26], Javascript coder Thomas Fuchs despairs,

> Nobody gives a s— anymore about anything. ¶ It's fine to post absolute tripe like this on a corporate website. ¶ It's fine to post quotes that were never said in a newspaper. ¶ It's fine to reference science that doesn't exist in scientific papers.... ¶ It's fine to steal, lie and scam people. ¶ All of this is hyped and cheered on by governments, organizations, businesses and individuals. ¶ What in the f—ing f— is wrong with everybody? [28]

And musing on what he finds to be the broken promises of AI-assisted coding, veteran software developer Mike Judge declares, "I'm not an angry person generally, but I can't stand what's happening to my industry." [37]

Clearly, these are not moderate sentiments, and neither are they run-of-the-mill gripes from employees under transient stress. In fact, bereavement, trauma, despair and anger are rapidly gaining gravity among practitioners' public statements, and such statements are often explicit about cause: Business models that lower quality and run roughshod over workers and users' interests; an over-hyped technology (generative AI) that speeds that deterioration; and putatively incompetent, immoral and/or authoritarian executives that enforce both. This is to say that practitioner distress follows the germination of polycrisis, from its technocapitalist origins through its acceleration via software.

Each of my interviewees to date has expressed some version of bereavement and / or betrayal by the industry. Moreover, their complaints continually fault the dictates of extractive technocapitalism:

> Most of the software apps, let's say consumer-facing, are built for addiction. And engagement, because that's how they measure software… performance, right? How long can you keep audience on your app? How much time they're spending on? These metrics are not humanized. — Interviewee 16, UX Designer

This designer echoes Seaver's observation that "Conflating [user] satisfaction and retention helped mediate a tension between developers, who often expressed to me a strong desire to help users, and business people, who wanted to capture them." [65, p. 431] They continue:

> I still care about user needs, because that's — that's almost like the integrity of my identity in my career. One of my strongest motivators is seeing how many s—y products out there, that are built by these egocentric stakeholders and these people with d— crowns on their head. — Interviewee 16, UX Designer

Nor are designers alone in attuning to capitalistic abuse of users:

> You can give people capabilities, but as the software company, you have, exactly, control over the limitations of those capabilities and their availability. And you can ask them to sign Terms of Services that say anything. Because it's not like I have a choice… We can't participate in life without agreeing to all the TOSs and the EULAs and everything. Um, so, it places a tremendous amount of power and control in large companies which are… very governed by the wider forces of what we call capitalism. Yeah, so, so that's a negative. And it's probably the defining negative. — Interviewee 4, Engineer

Many, like Sempere, take up Cory Doctorow's construct of "enshittification" [20] to describe the business models that have driven their work from user-friendly to user-hostile as they progressively claw back value to those closer to investors. Doctorow comments on how such erosion creates a form of "moral injury" for practitioners, whose employers had previously forborne to transgress their employees' social principles for fear of losing talent in tight labour markets but lost inhibition due to regulatory capture and monopolization. [22] Practitioners also cannily, if belatedly, recognize the increased market scarcity conditions that have followed the end of the US Federal Reserve's zero-interest rate policy: "The past decade and a half of the tech industry was a zero interest rate phenomenon," product manager Dare Obasanjo tweeted in 2023. [51] The "New Economy" seems to have faded to the old one, and practitioners find themselves discerning the ruins of a monument to their liberty on Greenspan's "new planet."

When asked about sources of practitioner dissatisfaction, interviewees are vociferous about linking their own morale to socially irresponsible corporate practices:

> The repeated narrative of technology [is] to democratise access and make things better (which it can! I've seen it!), yet we've also seen it concentrate power, amplify inequities, create new forms of harm. When you're working on systems that you know are extracting value rather than creating regenerative loops, how does it leave the ones involved? — Interviewee 14, Agile Coach

> Forty years sitting on top of a money printing machine. It doesn't matter how good your people are. It doesn't matter how good your code is, it doesn't matter whether you use statistical Markov chain generators to write code. And it doesn't matter whether anybody around you is happy. You just print more money. And that attitude has perked very, very deeply into our trade. And so, I think a lot of developers are unhappy because they have miserable lives and miserable jobs, and they've lost track of any sort of geek joy. — Interviewee 2, Developer Coach

A turning point in practitioners' experience of their own profession was Elon Musk's late 2022 takeover of Twitter and subsequent decimation of its workforce. Musk's confrontation with practitioners was almost caricatural: Those who did not commit within 24 hours to an "extremely hardcore" programme of "working long hours at high intensity" would be discarded. [73] Musk's layoffs provoked numerous condemnations about product quality and safety. [71, 75] But just as poignant for software professionals was to see the market conclude that Musk had not destroyed the company, and for a chain of other CEOs to interpret this conclusion as carte blanche to do likewise. [25] They now find themselves at a new level of precarity that remains to this day. Obasanjo comments on a February 2026 echo, at Block, of Musk's purge — this time tinged with the assertion that the jobs had been automated away:

> Jack Dorsey's layoffs at Block will give tech companies [permission to] make similar AI-driven workforce changes. Similar to how Elon's cuts at Twitter in 2022, kicked off a wave of layoffs and animosity towards tech employees that hasn't abated. [52]

In fact, Dorsey was at the time only the latest top executive to justify layoffs by invoking "AI"; the same had happened at Amazon [11], Autodesk [62], and Pinterest [32]. In the post-Twitter era, practitioners perceive generative AI tools in multiple ways — as a cudgel to be used against them in labour-hostile moves such as Dorsey's, as potentially powerful programming aids, as causes of ubiquitous code degradation and "slop" — in addition to echoing more general concerns about energy use, intellectual rights violation, deskilling, and psychosis. Many expressions of practitioner bereavement feature mention of AI tools, but to a degree this focus is symbolic: Practitioners often acknowledge what they dislike about AI is an acceleration or apotheosis of pre-existing pernicious trends. Thus software development consultant Baldur Bjarnason says,

> "Over the past year, every single time one of the apps or services I use suddenly became less reliable and more buggy, I never have to look far for the 'Claude [popular coding assistant LLM] is amazing and now writes most of my code' post for the devs involved.... This is happening IMO because of one of the fundamental issues with software dev (and this predates "AI"...): Most software projects fail and most of what gets shipped doesn't work. The way the industry is set up means there is little downside to shipping broken software." [13]

Mike Hoye, longtime systems administrator, comments, "We joke about vibe coding but we're at the tail end of 30 years of vibe management, vibe leadership and vibe capitalization in this industry and we have to solve the real problem." [33]

Practitioner Jürgen Güter, who blogs as tante, says,

> Software was doing bad before, standards of quality being largely nonexistent. But "AI" and the promise that you can just magically create software is pouring gasoline on the fire[.] [30]

Levels of emotion in such comments run high, as may be expected among those who believe their life's work to be going up in a figurative gasoline fire. But just as often, practitioners express their hurt more ironically, in mockery of their masters.

## 3 Setting Limits: Practitioner Mockery

On March 1, 2026, the social media account Fesshole, which collects anonymous confessions and publishes them online, posted: "Forced to use AI at work, hate it. Usage is monitored so I can't avoid it. I've set up mine to play a Sim City clone. It burns through tokens, making crap cities that quickly go bankrupt. Just been told that I'm getting a bonus for being a top user." [6] As the post's replies note, the veracity of the confession is uncertain, given the general traceability of digital activity at work [76, p. 10]. But more certain is the support the post drew from practitioners like software engineer Alice, who wrote "yup, this is the right way of dealing with employer forcing it - as opposed to pushing slop to unsuspecting open source projects" [2] Criticism among practitioners who commented focused on the externalities of burning tokens rather than the evasion of company policy.

This study finds no evidence for a new trend of actual sabotage among practitioners (although one interviewee reported subverting a leadership directive to fire a fixed percentage of their staff, and practitioners regularly share techniques for foiling LLMs and their associated scrapers online). Yet the derision inherent in such protestations bespeaks a new perception of power dynamics, far from the adulation of tech leaders common in dot-com days. As development consultant Bjarnason posts, 'In the absence of functioning regulation of the tech industry in the US... mockery and boycotts are the only mechanism available to us that have any hope of working.' [12] Boycotts from practitioners may not have brought down Big Tech as yet, but in its own way, mockery is trying.

Characterization of practitioners and executives as adversaries, with technical competence wholly on the practitioners' side, is a tradition in geek humour, as exemplified by 1981's "Putt's Law" — "Technology is dominated by two types of people, those who understand what they do not manage and those who manage what they do not understand" [60, p. 7] and 1995's "Dilbert Principle" — "Leadership is nature's way of removing morons from the productive flow." [1] This characterization has taken on a new and hostile edge, however, in this era when top tech executives control so much of what occurs not only inside their organizations but outside as well.

Oligarchic execs like Musk and Marc Andreessen are frequent targets for practitioner mockery, particularly in light of their outlandish public pronouncements. The latter's hifalutin self-heroization in his think-pieces — e.g. that his class are "conquering dragons and bringing home the spoils for our community" [5] — draws jeers from the likes of developer-turned-crypto-critic Molly White, who writes "[T]his is a man who built a web browser, not goddamn Beowulf" [74] and product management coaches Melissa and Jonathan Nightingale, who dub Andreessen's "It's Time to Build" [4] "The Poser Manifesto." [50] Per the Nightingales, Andreessen and his ilk are "very bad at actually building"; real builders are never "this boorish, this uncurious about where they might be wrong, this insistent that they are owed something." [50] Musk, for his part, is the main character of elonmusk.today, the remarkable site built by software engineer Ted Stein to document all of the failed promises Musk has made about his companies' technologies. They are voluminous, and Stein's use of plain fact as mockery has been effective enough to attract sabotage troubles of its own. [59]

Other ad hominem mockery directed from practitioners at oligarchs abounds [e.g. 45, 24, 61]. These figures make apt targets not only because they personify the wealth-inequality aspect of polycrisis — in a newly precarious economic era for practitioners — but also because they represent, via the idea of the “broligarchy” [79], tech’s descent into authoritarian alliance. In many cases their own unforced actions have set them up for easier ridicule — Peter Thiel’s preoccupations with the Antichrist and seasteadding, Mark Zuckerberg’s cage-match wrestling, Musk’s Nazi salute at Donald Trump’s second inauguration, Tim Cook’s presentation of a golden trophy to Trump, Dario Amodei’s doomsaying about Artificial General Intelligence, Satya Nadella’s claims of having CoPilot run his life. Short of cutting their power practically, such excesses provide practitioners ample opportunity to cut broligarchs down humorously.

Beyond targeting particular oligarchs, practitioners frequently glory in mocking their type of leadership. Embedded systems programmer Jayne comments in agreement with Mike Hoye’s post quoted above about “vibe management”: “The mirrortocracy in silicon valley fundraising is so bad, it's like a s—y clone wars remake sponsored by Patagonia[.]” [35] Kevin Beaumont makes his irreverence for technical leadership specific: “Things I’ve been asked to produce over the years as a manager in security: ¶ - Blockchain strategy ¶ - Metaverse strategy ¶ - Generative AI strategy ¶ Amount of times this has had any value: 0.” [7] And data scientist Nikhil Suresh (who blogs as Ludicity) has risen to blogospheric prominence for his skewering of retrograde tech leadership, e.g. “That level of self-regard is narcissistic to such a degree that I have screeched so far past horror that I've looped around to arrive at a grudging respect.” [70]

Humour can be a substitute for more aggressive confrontation, or it can be a prelude. One clue is the relative directness or indirectness of the argument implicit in the joke. Little mirth is to be detected, for example, in a meme posted by the tech-humour account Fake Scrum Stats Memes and Humor on March 12, 2026 (figure 1). Using stills from the series *Star Trek: Deep Space Nine*, the meme shows one frame with a young girl embracing a uniformed man, with the caption “How management wants the employees to treat them,” and a second frame in which the man tosses the girl aside, with the caption “How they treat the employees.” This setup echoes many propaganda posters showing enemy classes violently mistreating innocents, such as Viktor Deni's 1919 poster “Death to Capitalism or Death Under Capitalism” (figure 2) or Harry Ryle Hopps’ “Destroy This Mad Brute: Enlist” (figure 3), with only a touch of irony to leaven the antagonism.

But in the post-Twitter era, practitioners are often dispensing with levity altogether to frankly endorse anticapitalist movements and ideologies that they find fit for the struggle indicated by such imagery.

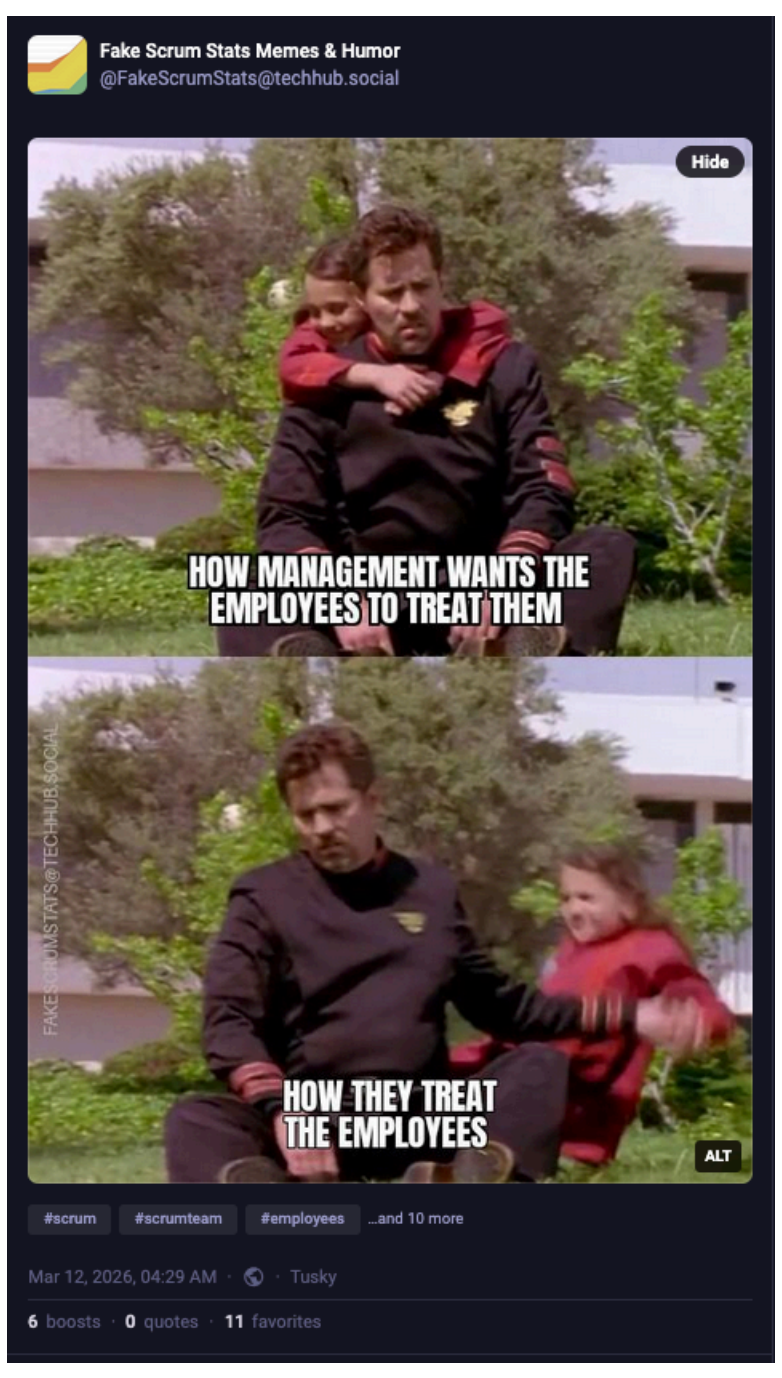

Figure 1: **Meme posted to Mastodon by the account Fake Scrum Stats Memes & Humor March 12, 2026**

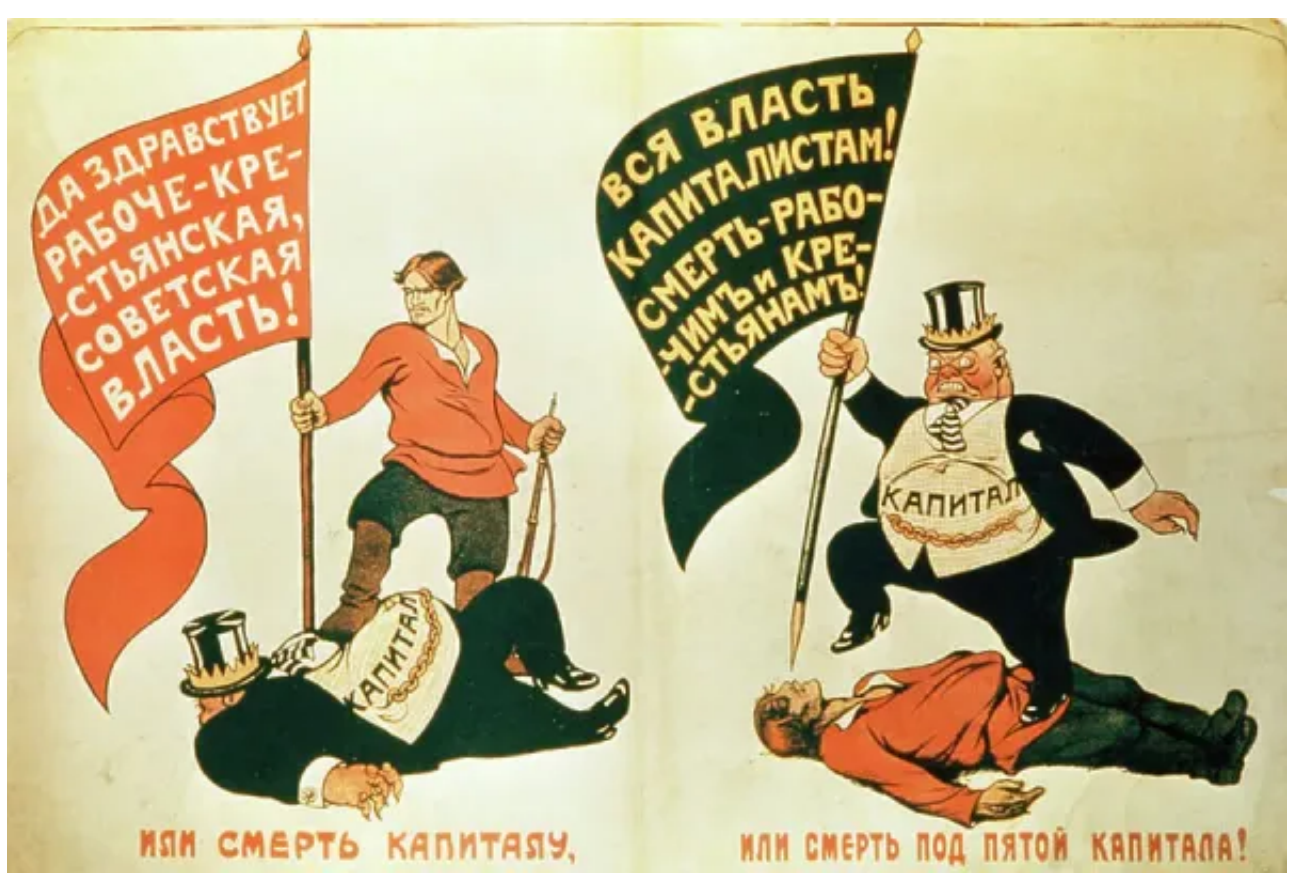

Figure 2: **Viktor Deni, “Death to Capitalism or Death Under Capitalism,” 1919**

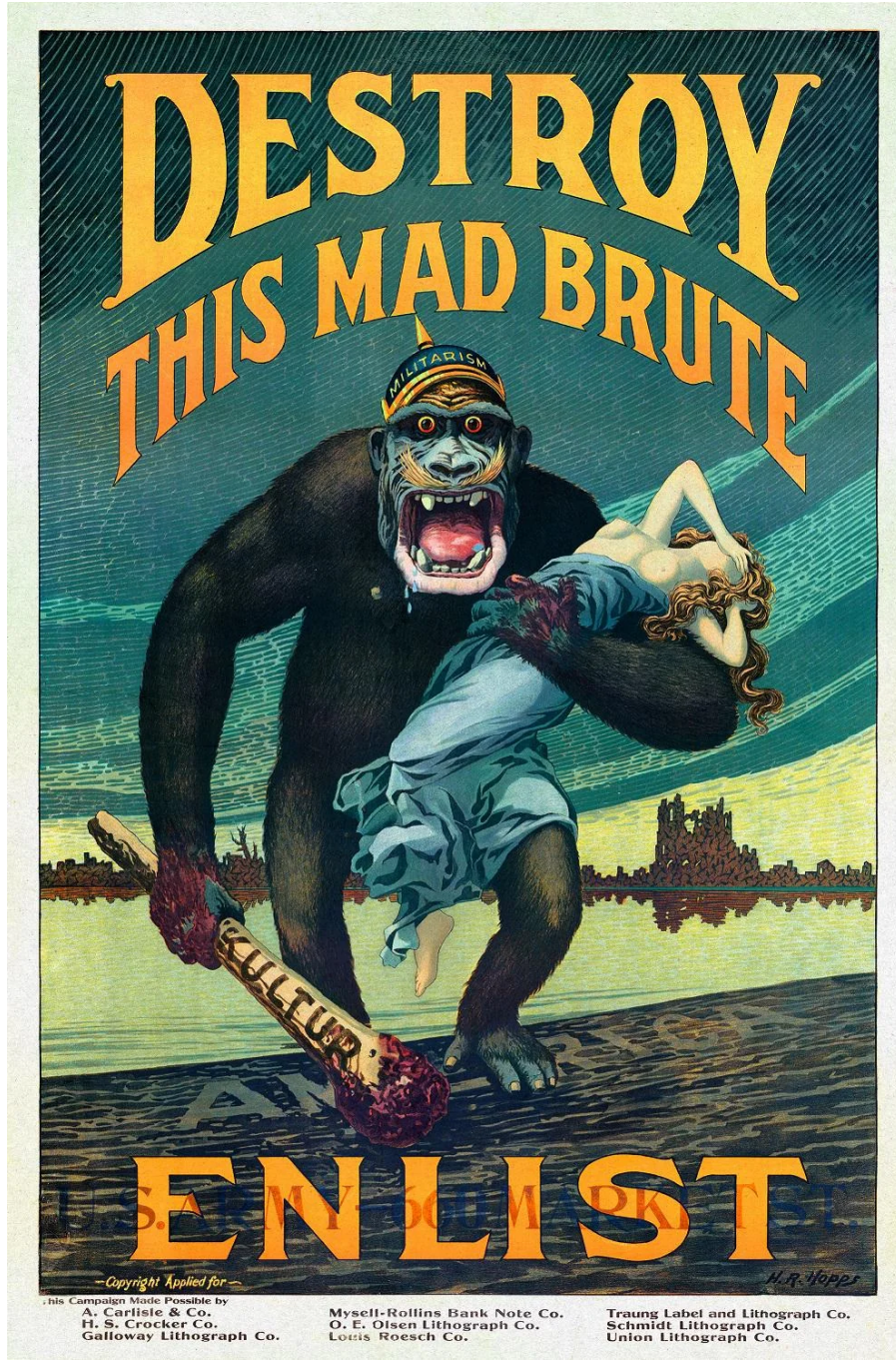

Figure 1: **Harry Ryle Hopps, "Destroy This Mad Brute: Enlist," 1917**

## 3 Setting Limits: Practitioner Politics

Listening to UX Designers and Data Scientists, Robert Dorschel finds evidence of "a new middle-class identity among contemporary tech workers." [23, p. 4] This class identity, he grants, comes with a caveat: "[W]hile tech workers identify with ideals of social justice and critiques of power structures, there is very little evidence of these commitments being enacted in tangible, impactful ways." [23, p. 114] But if software practitioners are not actively storming the Bastille, they are seeking and naming alternative ideologies opposed to technocapitalism with energy and frequency. Among these are communism, anarchism, labour collectivism, and Luddism.

Interviewee partner Clare Sudbery, a Principal Developer who espouses the Extreme Programming (XP) variety of agile software practice, recalls Trotsky:

> A minority of teams... are really fully following XP coding practices. Why is that? Well, I think it's because you can't have socialism in one country. I think it's because... as long as capitalism is still the reigning system, any attempt you make to make these changes whilst pretending that you're not actually fundamentally trying to overthrow capitalism [means] It's still gonna win! Because, like, capitalism is still there... We're pretending that we're just changing software practices. I think really what we're trying to do is overthrow capitalism. — Clare Sudbery, Principal Developer

XP is an approach to programming that emphasizes aspects of care — careful and continuing testing, collegial pair-programming, routine refactoring in the interests of sustainability. [36] As such, it is "[s]ometimes... absolutely contrary to accepted wisdom" [8] but it is still overwhelmingly practiced within conventional, capitalist software organizations. Sudbery's comments not only endorse overthrow of technocapitalism; they also naturalize such overthrow in tech practice: Contestation of capitalism is *entailed* in XP.

Interviewee 15, a systems administrator, also uses the vocabulary of early Soviet Communism: "You're familiar with Lenin's phrase of a useful idiot, right? It's somebody who is... you know, a devoted advocate for a cause they don't actually understand." They propose that software practitioners have long been in this position because "it was very politically useful for us to be able to go 30 years in this industry with people believing that they're unique and special, instead of forming a union."

Paralleling Sudbery's connection between software practice and anti-capitalism, Clojure programmer "Chip Butty" remarks: "If it isn't an anarcho-syndicalist system, then it isn't #agile ¶ If it isn't anarcho-communism then it isn't #FreeSoftware" [14]

Similarly, programmer grmpyprogrammer writes, "The pinko commie in me thinks it's wild that a group of people who think they are rebellious visionaries are adopting the tools of their oppressors and working hard to turn themselves into easily-replaced cogs in a machine." [29]

It appears practitioners' rising class consciousness is accompanied by a *self*-consciousness regarding their former general disinterest. Game developer JP LeBreton summarizes,

> a theory on the total demobilization of programmers (in aggregate, not you specifically) as a class:
> - comfortable life (salary, benefits, retirement plan)
> - never protested or resisted anything / did but it didn't seem to do much
> - never organized their workplace / community
> - computers are a solace from messy real world stuff
> - suddenly computers are subject to this new political economy
> - feel totally helpless and concede to what's happening right now as inevitable [39]

Unionization, mentioned by LeBreton and by Interviewee 15, is the most institutional labour response to employer excesses among those mentioned by interviewees and online sources. Collectivization among tech workers is surely greater than it was. Software designer Ethan Marcotte, in his 2023 *You Deserve A Tech Union*, documents milestones reached in the 2010s such as the 2017 "Never Again" pledge not to build databases that would be used to deprive people's rights, Google employees' protests against Project Maven in 2018, and GitHub workers' 2019 anti-ICE petition. [41] In 2021, workers at Google's parent company, Alphabet, unionized. The web site Collective Action in Tech, which says it "attempts to document all collective action from workers in the tech industry," lists 569 strikes, open letters, protests, and other actions. As this paper underwent revision, IT employees at the University of California formed the largest American tech worker union yet. [44]

Nevertheless, unionization in the tech sector still has a long way to go. The US Bureau of Labor Statistics reports that 3.7% of salary and wage workers in "Computer and Mathematical Occupations" were union members in 2024, essentially unchanged from 2020. [80] While Dorschel finds that tech workers "express sympathy for unions," he also finds their "individualist and deeply incorporated modus operandi can be considered largely responsible" for relatively subdued responses to date to layoffs and tech entanglement with authoritarianism. [23, pp. 16, 114] Despite a continuing string of impressive labour achievements in the software practice space — e.g. the 5000+ signatures collected by Amazon Employees for Climate Justice for their late 2025 open letter demanding the company respect climate, labour, and civil justice goals in its pursuit of AI development [3] — collectivization remains a reality for a small minority of practitioners.

If communism and unionism are the most conspicuous of anticapitalist ideologies broached by practitioners, they are not the only ones. Brian Marick, a signatory of the original Agile Manifesto [9], offered an update in 2009 called "AR⊗TA" [42] in which he explicitly embraces Anarcho-syndicalism, as does "Butty," quoted above, opining that Agile as a movement had lost its original anarchism.

Anarchism may be seen as a fairly malleable concept for practitioners that resists the kind of strict formalization that might produce contra-indications for endorsement, while yet providing enough power to focus reformative energy. Another such ideology that has nevertheless gained in definition as well as popularity, via its rehabilitation among practitioners, is Luddism.

Luddism, a movement that named itself for a mythic misbehaving apprentice, conducted an actual rebellion against business owners' implementation of technology and associated policies, peaking early in the 19th century in England, and acquired more than a little mythology of its own, has long been popularly conceived as a movement against technological progress, peopled by resentful and reactionary technophobes. But this conception is clearly changing: In 2021 Mueller found "a palpable rise in Luddite sentiments, as well as anti-capitalist ones" among workers generally. [46, p. 8] He also traces a specific Luddite tradition in digital technology, claiming, "Time after time, skilled computer users have mobilized to protect their established autonomous ways of crafting, exchanging, and using digital artifacts, rather than behaving as corporations… would have wished." [46, p. 108]

In their defiant and sometimes admonitory use of a term that has for long had, and still retains, pejorative connotation, practitioners who endorse Luddism and / or identify as Luddites are engaging in an act of linguistic *reclamation*, defined by Popa-Wyatt as "a form of socio-political protest that seeks to re-shape oppressive social practices by controlling what can be done with words." [57] In the case of Luddism, which itself denotes a protesting group, reclamation is protest on top of protest.

Mueller draws a line between Luddism and Marxism: "[T]o be a good Marxist is to also be a Luddite." [46, p. 5] However, despite sometimes separately endorsing Communism as noted above, I did not find software practitioners making this connection themselves. Instead, *Luddism* and *Luddites* seem to be used symbolically in the resistant rhetorics of software practitioners. Often the terms are attached via hashtag without other direct reference, apparently to invoke a larger movement or tradition of resistance without explicitly arguing the movement's merits. Thus developer Chris Pitts responds to news of Microsoft's internal memo stating that "Using AI is no longer optional" [69] with this post:

> Ooooh boy…I do hope I get one of these ridiculous edicts.
> "Why did this production problem happen? Why doesn't this code work?"
> "It wasn't me. It was the AI. I used it exactly as you requested"
> <phones Union lawyers, just in case>
> #Luddites [56]

Technical writer Miguel Alfonso Caetano appends the hashtag #Neoluddism to his posted quotation [15] of an article critiquing AI through Lewis Mumford's "myth of the machine" construct. [43] Likewise, tante (Jürgen Gueter), writes, "I am a luddite and to me there is a lot of joy in technology." [31] Gueter's blog is titled "SMASHING FRAMES," an apparent pun on the historical Luddites' resistance method and the reconfiguration of cognitive / interpretive frames.

In attempt to contextualize online use of the Luddite label, I used the Zeeschuimer browser plugin with a new account on the business networking site LinkedIn to search for the hashtag #Luddite and scrape the resulting posts into a CSV table. For each result, I manually determined (via bio and experience details on the site) whether the poster was or was not a software practitioner. I also determined (from the text of each post and contextual clues) whether its use of #Luddite was critical, uncritical or indeterminate. Uncritical examples included pejorative uses and protestations of disidentification; critical ones included those emphasizing historical recontextualization or evident critique of dominant tech practices. Because the site's search algorithm apparently truncated results to 145 total (2 from 2020, 1 from 2023, 4 from 2024, 116 from 2025, and 22 from the partial year 2026 (January 1 through collection on March 2015), results are not to be considered dispositive. However, one striking result is that of the 6 practitioner-posted uses of the hashtag shown by the algorithm so far in 2026, 5 were critical and 1 indeterminate, as compared to 9 critical, 6 uncritical and 0 indeterminate in 2025.

(Note that public API access to data from more specifically practitioner-oriented communities, such as subreddits on reddit.com, might have yielded more informative results, but such access has been curtailed over the past several years. This curtailment itself is often associated with power-consolidating moves on the part of tech executives [49], putting practitioners and researchers in a similar position.)

Mueller is hopeful for the rise of Luddism as a boost to broader movements toward economic justice and sustainability: "Luddism might articulate with interest in the politics of degrowth." [46, p. 132] However, one salient point for the Computing Within Limits community is that, thus far, no interviewee and very few sources of online public data for this study have spontaneously mentioned Decomputing, Degrowth, or related frameworks. This is not to say that they oppose such frameworks' principles, but rather that, in comparison to the movements and ideologies mentioned above, they remain unfamiliar. While the emotions that they channel into the forms of expression I analyze in this paper are consistent with literature that describes and predicts transition away from dominant unsustainable systems [19, 54], they do not appear to evidence intentional, organized participation in such a transition.

## 3 Conclusion

Bereavement, mockery, and political noisemaking do not themselves add up to transformative change. At the same time, transformative change is unlikely to occur without them. Necessary if insufficient conditions, then, exist within the class of software practitioners, if that class hopes to help establish either an effective resistance to the socio-economic paradigm that causes their current distress or a plausible replacement for it.

Dorschel calls this a "*hinge moment* in capitalism," when the tech workers he studies are conceiving commonality with others in polycrisis, positing "there is, perhaps, something in the air, something up for grabs." [23, p. 178] This study confirms: Software practitioners are in crisis — of authoritarianism, of enshittification, of precarity, of social and planetary endangerment — and are casting about emotionally, ironically, and ideologically for ways forward.

But this same moment has seen effective inhibition of such potential by the forces of technocapital which are, far from ceding the fight, stepping it up. In the time since Dorschel's optimism was expressed, tech owners' ravishment of the job market for practitioners has intensified, with practical results. Steven Levy, longtime editor of *Wired*, comments: "It used to be that when tech's leaders fell short of their lofty values, employees kept them honest… Implicit was the threat that the activists could easily find jobs elsewhere." [40] No longer.

Will the elimination of relatively indirect routes for practitioners' resistance permanently damage their ability to resist at all, or will it galvanize them to find ways to act more effectively at scale? Unionization is often posed as an answer to this dilemma [e.g. 18, 21], but multiple barriers to this option exist, as discussed above, to say nothing of traditional union-busting tactics available to employers — and membership numbers have stagnated.

At the same time, there is little likelihood of practitioners' solving the polycrisis each on their own. Other forms of collectivization and organization exist besides unionism: Interest groups such as the Naysayers Club [48], Prosocial Tech Collab [58], and the Labor Tech Research Network [38] are connecting likeminded practitioners and scholars in critique and collaborative revisioning. Additional scholarly, tech-professional, and activist attention should be paid to such groups and their potential for cross-collaboration to channel distress and critique into effective change. In many cases, such efforts are proudly small-scale and local, to differentiate themselves from the hyperscaling valourized in platform capitalism. Still, common organization will probably be needed to confront common crisis. In this moment, software practitioners possess the requisite perspicacity and urgency to be persuaded into such common organization.

## ACKNOWLEDGMENTS

I would like to express intense gratitude to all of the interviewee partners who have participated in my dissertation study. Several who chose to waive anonymity have nevertheless been left anonymous in this paper in an abundance of caution, since it precedes the final dissertation product, and I look forward to thanking them by name soon! In addition, Hana Holubec offered very valuable insight and feedback about practitioner mockery. Thanks also to my wonderful supervisor, Dr. Robert Gehl, for his support.